\documentclass[aps,prl,amssymb,showpacs,twocolumn]{revtex4-1}

\newcommand{\bA}{{\bf A}}

\newcommand{\beq}{\begin{equation}}
\newcommand{\beqn}{\begin{eqnarray}}
\newcommand{\eeq}{\end{equation}}
\newcommand{\eeqn}{\end{eqnarray}}

\newcommand{\vect}[1]{\mathbf{#1}}
\newcommand{\ket}[1]{\vert #1 \rangle}
\newcommand{\bra}[1]{\langle #1 \vert}

\makeatletter

\newcommand{\Rmnum}[1]{\expandafter\@slowromancap\romannumeral #1@}
\makeatother

\usepackage{graphics,graphicx,amsmath}

\usepackage{tabularx,float}
\usepackage{epsfig}

\usepackage{subfigure}

\usepackage{bm}
\usepackage{wasysym}

\usepackage{bm}

\usepackage{color}
\usepackage{verbatim}

\begin{document}

\date{\today}

\author{X. Lu and M. O. Goerbig}
\affiliation{Laboratoire de Physique des Solides, Univ. Paris-Sud, Universit\'e Paris Saclay, CNRS, UMR 8502, F-91405 Orsay Cedex, France}

\title{Magneto-optical Signatures of Volkov-Pankratov States in Topological Insulators}

\begin{abstract}

In addition to the usual chiral surface states, massive surface states can arise at a smooth interface between a topological and a trivial bulk insulator. While not subject to topological protection as the chiral states, these massive states, theorized by Volkov and Pankratov in the 1980s, reflect nevertheless emergent Dirac physics at the interface. We study theoretically the magneto-optical response of these surface states, which is strikingly different from that of the bulk states. Most saliently, we show that these states can be identified clearly in the presence of a magnetic field and its orientation with respect to the interface.

\end{abstract}


\maketitle

The bulk-edge correspondence is certainly one of the most spectacular properties of topological materials. It states that a topological material, described by a topological invariant, necessarily displays a chiral state at the surface (for a three-dimensional material) or at the edge (in the case of a two-dimensional system) when they are put in contact with a trivial insulator, e.g. the vacuum, with zero topological invariant \cite{z2,reviewtopo}. These states have indeed been observed experimentally in angular-resolved photo-emission spectroscopy \cite{exp1,exp2,exp3} or via quantized conductances, e.g. in HgTe/CdTe quantum wells \cite{konig2007}. Already back in the 1980s, well before the advent of topological band theory, it was realized that the gap inversion over an interface necessarily implies a gap closing at the interface with emergent relativistic $(D-1)$-dimensional carriers for $D$-dimensional bulk materials \cite{volkov1985two,pankratov1987supersymmetry,fradkin1986physical,boyanovsky1987anomalous}. It was later understood that the gap inversion was a necessary consequence of the change in the quantized topological bulk invariant. 

In addition to the emergence of a massless (ultra-)relativistic chiral state, Volkov and Pankratov showed that additional massive states can arise at the $(D-1)$-dimensional surface or interface \cite{pankratov1987supersymmetry}. This requires that the gap changes sufficiently smoothly over a distance $\ell$ that is large as compared to an intrinsic length $\xi=\hbar v/\Delta_s$, where $\Delta_s$ is half of the bulk gap and $v$ a velocity that characterizes the bulk dispersion relation. 
Indeed, the number of visible surface states beyond the chiral one is given by the ratio $\ell/\xi$, which characterizes a topological heterojunction (THJ) \cite{tchoumakov2017volkov}. First experimental evidence of these massive Volkov-Pankratov (VP) states stems from transport measurements on the surface of HgTe systems under a strong electric field -- oscillations in the conductivity indicate the population of VP states that serve as additional scattering channels \cite{inhofer2017observation}. 

Beyond these sophisticated transport measurements, a direct evidence of VP states apart from the chiral ones is yet lacking. A promising technique for their identification would be (magneto-)optical spectroscopy that has been shown to be a valuable technique in the identification of topological and relativistic materials in the past \cite{sadowski,orlita2011,orlita2015,crassee2017,hakl2018,krizman2018}. The theoretical analysis of such spectroscopic signatures of VP states of three-dimensional (3D) topological insulators (TIs) is the subject of the present Letter. We show here that VP states can be probed spectroscopically and derive their selection rules both in the absence and the presence of a magnetic field. The first clear-cut signature naturally consists of transition lines corresponding to energies below the bulk gap. Furthermore, the dimensionality of the VP states, as opposed to bulk states, manifests itself in a specific shape of the spectroscopic absorption lines. Most saliently, we find that the transition lines associated with VP states show a particular evolution under the application of a magnetic field and its orientation with respect to the interface or surface. 

While the emergence of chiral and massive VP states is a general feature, we exemplify them here with the help of a variant of a Hamiltonian that was originally proposed by Zhang \textit{et al.} in the study of Bi$_2$Se$_3$ \cite{zhang2012},
\begin{equation}
        \label{eq:h03dti}
        H_0 = \Delta(z) \tau_z + \hbar v k_z \tau_y + \hbar v \tau_x (k_y \sigma_x - k_x \sigma_y),
\end{equation}
where $\tau_{\mu}$ and $\sigma_{\nu}$ are Pauli matrices representing the orbital and spin degrees of freedom, respectively. The interface between the 3D TI and a trivial insulator is modeled by the gap parameter $\Delta (z)$, which varies between a negative value (for $z<0$) and a positive one (for $z>0$) in the $z$-direction that we choose perpendicular to the interface. While a natural functional form of the varying gap parameter is the function $\tanh(z)$ \cite{pankratov1987supersymmetry}, we use here a simplified linear form, $\Delta(z)=\Delta_s z/\ell$, that yields the same qualitative spectra, as it was shown in Ref. \cite{tchoumakov2017volkov}. One notices from the form of Hamiltonian (\ref{eq:h03dti}) that its spectrum consists of two-dimensional (2D) bands $E_{\lambda,n}(k_x,k_y)$ that disperse in $k_x$ and $k_y$ while the motion in the $z$-direction is quantized due to the quantum-mechanical noncommutativity between $z$ and $k_z$. 
These bands are readily obtained with the help of a unitary transformation $T=\exp(i\pi\tau_y/4)$, which interchanges  the Pauli matrices $\tau_x$ and $\tau_z$. As a consequence the Hamiltonian (\ref{eq:h03dti}) can be rewritten as 
      \begin{gather}
    	\label{eq:hc3d}
     	H_T = T H T^{\dagger} = \hbar v 
        \begin{bmatrix}
        	0 & k_+ & \frac{\sqrt{2}}{l_s} \hat{c} & 0   \\
            k_- & 0 & 0 & \frac{\sqrt{2}}{l_s} \hat{c}   \\
            \frac{\sqrt{2}}{l_s} \hat{c}^{\dagger} & 0 & 0 & -k_+ \\
            0 & \frac{\sqrt{2}}{l_s} \hat{c}^{\dagger} & -k_- & 0 
        \end{bmatrix},
    \end{gather}
where $k_\pm =k_y \pm i k_x$ in complex notation, and we have defined the ladder operators 
\begin{equation} \label{eq:c3d}
        \hat{c} = - \frac{l_s}{\sqrt{2}} \left( \frac{z}{l^2_s} + i k_z \right) ~~ \text{and} ~~
                \hat{c}^{\dagger} = - \frac{l_s}{\sqrt{2}} \left( \frac{z}{l^2_s} - i k_z \right),
\end{equation}
which satisfy the usual commutation relations $[c,c^{\dagger}]=1$. The length $l_s=\sqrt{\ell\xi}$ plays the role of a magnetic length and characterizes the spatial extension of the surface-state wavefunctions in the $z$ direction. 
The Hamiltonian is now easily diagonalized, and one obtains the usual chiral bands 
\begin{equation}\label{eq:chiral}
     E_{\lambda,n=0}(k_\parallel)  = \lambda \hbar v k_\parallel,
\end{equation}
with $k_\parallel=\sqrt{k_x^2+k_y^2}$, $\lambda=\pm$, and the massive VP states 
\begin{equation}\label{eq:VP}
         E_{\lambda,n}(k_\parallel)  = \lambda \hbar v \sqrt{k_\parallel^2 + \frac{2n}{l_s^2}},
\end{equation}
which are twofold degenerate. One notices that the dispersion of the chiral states depends only on bulk parameters and survives even in the limit of a sharp interface $\ell\ll \xi$, in agreement with the Jackiw-Rebbi argument \cite{JR1976}, while the massive VP bands are gapped with a minimal gap of $\hbar\sqrt{2n} v/l_s$ at $k_\parallel=0$. In order for these VP states to be well defined and visible, one needs to require that the minimal gap be smaller than the bulk gap $2\Delta_s$. This yields the condition $n\leq \ell/\xi$, in agreement with the argument that the spatial extent of the wavefunction in the $z$-direction (the ``cyclotron radius'' $\sqrt{2n}l_s$) needs to be smaller than the width of the interface, $\sqrt{2n}l_s\leq \ell$.

Before discussing the optical properties of these states, let us consider the effect of a magnetic field. Naturally, a magnetic field applied perpendicular to the interface (in the $z$-direction) quantizes the 2D surface states into Landau levels -- one needs to replace $k_y\pm i k_x = \sqrt{2}b^{(\dagger)}/l_{B_\perp}$,  where $b$ and $b^{\dagger}$ are a second set of ladder operators, $l_{B_\perp}=\sqrt{\hbar/eB_\perp}$ is the magnetic length associated with the true magnetic field $B_\perp$, and one obtains the spectrum 
\begin{equation}\label{eq:LLs}
  E_{\lambda,n,m} = \lambda \hbar v \sqrt{\frac{2n}{l_s^2} + \frac{2m}{l_{B_\perp}^2}}.
\end{equation}
In contrast, if the magnetic field $B_\parallel$ is applied in a direction parallel to the interface, say the $x$-direction, it conspires with the confinement in the THJ. By choosing the Landau gauge $\bA=(0,-B_\parallel z,0)$ and thus using the Peierls substitution $k_x\rightarrow k_x$ and $k_y\rightarrow k_y -eB_\parallel z/\hbar$, one can again transform the Hamiltonian from the form (\ref{eq:hc3d}), but now with the help of the additional unitary transformation $T = \exp \left( -i \theta \tau_y \sigma_x / 2  \right)$, in terms of the angle $\theta$ defined by $\cos\theta= \gamma^2/l_s^2$ (or $\sin\theta=\gamma^2/l_B^2$), where $\gamma$ plays the role of an effective magnetic length composed of the true magnetic length and $l_s$, by $\gamma^{-4}=l_s^{-4}+l_{B_\parallel}^{-4}$. One thus obtains for the modified surface bands\textbf{ [SM1]}
\begin{equation}\label{eq:chiral2}
  E_{\lambda,n=0}  (k_x,k_y)= \lambda \hbar v k_{\parallel,\theta}
\end{equation}
in the chiral ($n=0$) case and
\begin{equation}
  E_{\lambda,n} (k_x,k_y) = \lambda \hbar v \sqrt{k_{\parallel,\theta}^2 + \frac{2n}{\gamma^2}} 
\end{equation}
for $n\neq 0$. In addition to the effective magnetic length, the in-plane magnetic field renders the in-plane wave vector anisotropic, $k_{\parallel,\theta}^2 = k_x^2 + k_y^2 \cos^2 \theta$. As compared to the case without a magnetic field, one notices that the spacing between the surface bands is enhanced by a factor of $(1+l_s^4/l_{B_\parallel}^4)^{1/4}$ and the dispersion in the $y$-direction is flattened by a factor of $\cos\theta$. In other words, since $\gamma$ is always smaller than $l_s$ when the magnetic field is applied, the surface becomes ``sharper". Most saliently, one notices that both the case of a perpendicular magnetic field, where one obtains a fully quantized Landau-level spectrum, and that of an in-plane field, where one maintains 2D surface bands, are strikingly different from the spectrum of the bulk bands, which are effectively one-dimensional (1D) in the presence of a magnetic field, with 
\begin{equation}\label{eq:bulkbands}
 E^{\text{bulk}}_{\pm, n} (\vect{k}) = \pm \hbar v \sqrt{k_z^2 + \xi^{-2}+ \frac{2n}{l_B^2}}.
\end{equation}
As we show below, the difference in the effective dimensionality of the bands has a clear signature in the shape of the optical transition lines, which is determined by the (joint) density of states (DOS). 

In (magneto-)optical spectroscopy, one has, both in the direct transmission spectra and in reflectivity, access to the optical conductivity at frequency $\omega$, which can be calculated via the Kubo formula
\begin{eqnarray}
 \label{eq:kubooc3d}
 \nonumber
        \sigma_{ij} (\omega) &=& i \hbar e^2 \sum_{\substack{m,n \in \mathbb{N}\\ \lambda,\lambda' }} \iint_{- \infty}^{+\infty} 
        \frac{ d k_x d k_y}{4\pi^2} \frac{f_D (E_{\lambda,n}) - f_D (E_{\lambda',m})} {E_{\lambda',m} - E_{\lambda,n} - \hbar \omega + i \delta}\\ 
        && \times 
        \frac{ \bra{\psi^{\lambda'}_m}\hat{v}_i \ket{\psi^{\lambda}_n}  \bra{\psi^{\lambda'}_m}\hat{v}_j\ket{\psi^{\lambda}_n}^* }{E_{\lambda',m} - E_{\lambda,n}}.
\end{eqnarray}
Here, $f_D(E)$ is the Fermi-Dirac distribution function, and the parameter $\delta$ takes into account, on the phenomenological level, a possible level broadening due to disorder. Notice that in the case of fully quantized Landau levels, one needs to replace the double integral over the wave vector components $k_x$ and $k_y$ (divided by $4\pi^2$) by a factor of $n_B=1/2\pi l_{B_\perp}^2$. The matrix elements $\bra{\psi^{\lambda'}_m}\hat{v}_i \ket{\psi^{\lambda}_n}$ of the $i$-th component of the velocity operator are evaluated in the eigenstates $|\psi^\lambda_n\rangle$ of the Hamiltonian and encode the optical selection rules as a function of the polarization of the radiation field. In the absence of a magnetic field, light with a polarization in the $x$- or $y$-direction couples surface bands with the same band index $n\rightarrow n$, i.e. only transitions involving different $\lambda$ are visible, while light polarized along the $z$-direction yields transitions $n\rightarrow n\pm 1$, regardless of the value of $\lambda$. A magnetic field along the $x$-direction does not alter the selection rules for $x-$ and $z$-polarized light, while $y$-polarized light now also induces transitions $n\rightarrow n\pm 1$. This is a direct consequence of the unitary transformation, which we used above and that mixes the $y-$ and $z-$ components for $y-$polarized light. If a magnetic field is applied perpendicular to the interface, one finds the following selection rules: for a polarization in the $z-$direction, one has $n\rightarrow n\pm 1$ for the quantum number associated with the confinement, while light with this polarization does not change the Landau-level quantum number, $m\rightarrow m$. On the contrary, one has the opposite selection rules for light polarized in the $xy$ plane, with $n\rightarrow n$ and $m\rightarrow m\pm 1$.

        	
            	\begin{figure}[htpb]
                \centering
                \subfigure{(a)\label{fig:3dsigmazzBx}
                \includegraphics[width= 0.4 \textwidth]{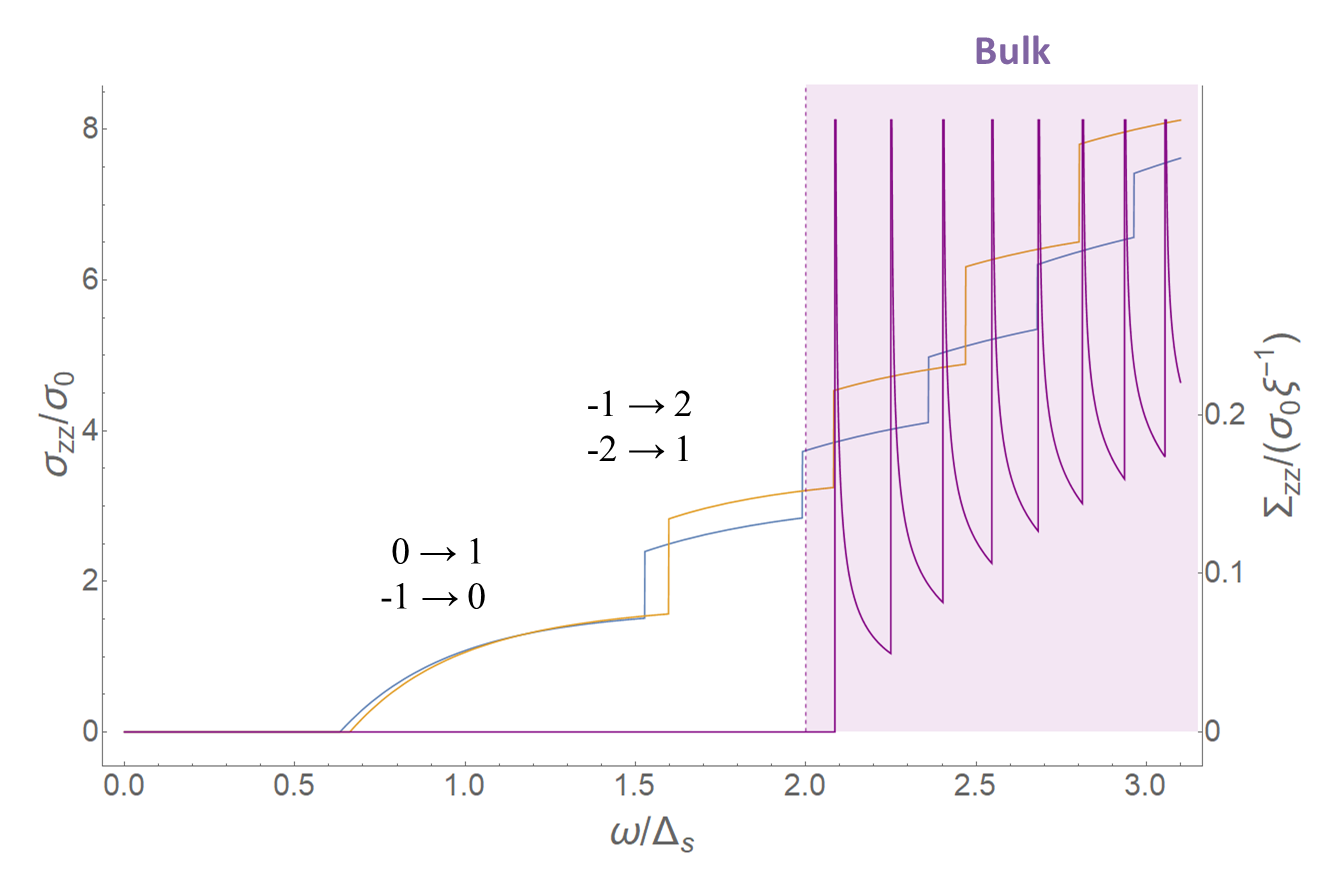}
                }
                \subfigure{(b)\label{fig:3dsigmaxxBx}
                \includegraphics[width= 0.4 \textwidth]{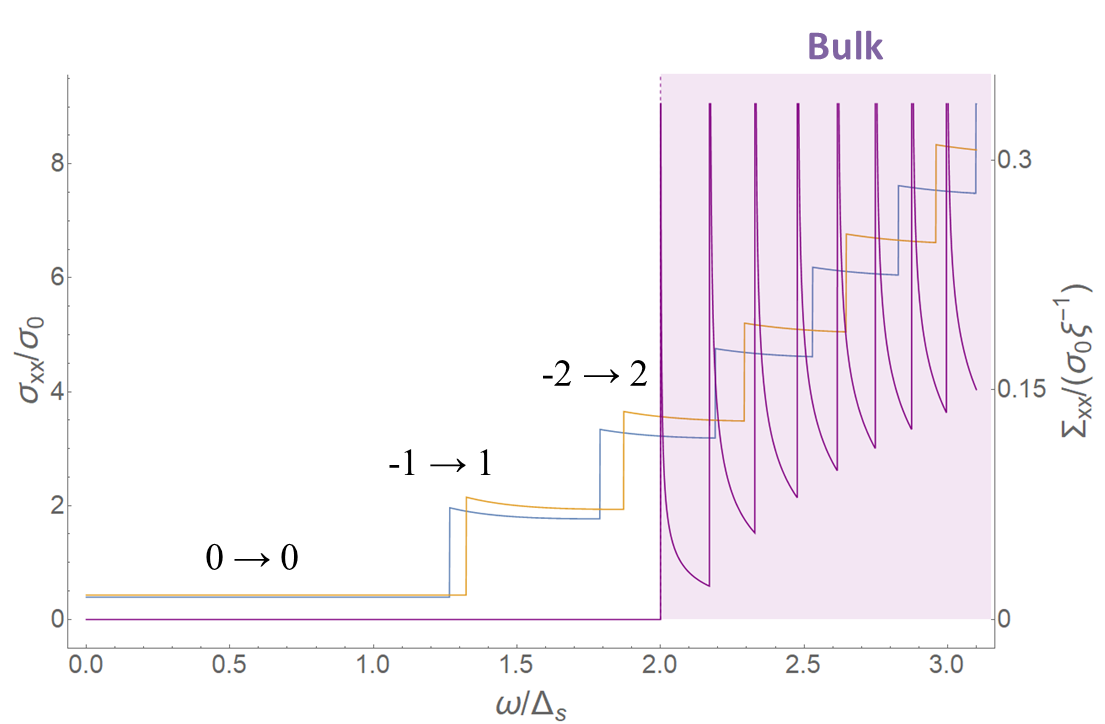}
                }
                \subfigure{(c)\label{fig:3dsigmayyBx}
                \includegraphics[width= 0.4 \textwidth]{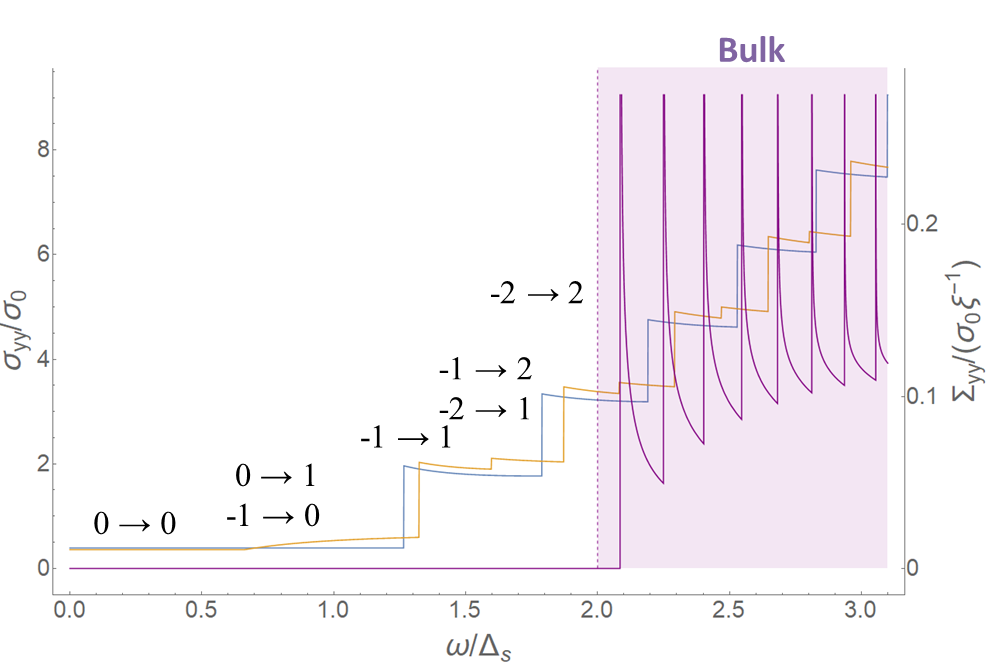}
                }
                \caption{Optical conductivities
                $\Re[\sigma_{zz}]$, $\Re[\sigma_{xx}]$ and $\Re[\sigma_{yy}]$, in units of $\sigma_0 = e^2/h$,
                in the absence (blue) and the presence (orange) of a 
                magnetic field parallel to the interface. The magnetic field is chosen such that $l_B=1.5 l_s$,  and the excitation energy $\hbar\omega$ is given in 
                units of the bulk gap parameter $\Delta_s$. The 
                corresponding optical transitions of each peak are labeled above. The light-purple shaded area indicates the transition threshold for excitations in the 
                bulk gap (purple lines), where surface-state transitions should no longer be observable. }
                \label{fig:3dsigmaBx}
                \end{figure}
                

The selection rules obtained above are illustrated in the (real part of the) optical conductivity (\ref{eq:kubooc3d}), where we have fixed, in our calculations \textbf{[SM2]}, an interface width of $\ell=5\xi$. All the calculations are done for half-filled systems at zero temperature. Let us first consider the optical conductivity $\Re[\sigma_{jj}]$ for the three polarizations $j=x,y$, and $z$, in the absence of a magnetic field. They are plotted in blue in Fig. \ref{fig:3dsigmaBx}(a) - (c). Apart from the selection rules, one notices that the surface-state transitions have a clearly distinct line shape as compared to bulk transitions. Generally speaking, the shape of the transition lines in the optical conductivity is given by the ratio between the scaling in frequency of the joint DOS and the frequency itself, which plays the role of an inverse scattering time, 
\begin{equation}
 \sigma(\omega) \sim \frac{\text{JDOS}}{\omega}.
\end{equation}
As mentioned above, the surface-state transitions involve effectively 2D bands -- while the DOS of the chiral surface states is linear in $\omega$ due to the linear dispersion relation, that associated with the massive VP states is constant due to their parabolic dispersion at the band bottom. Transitions involving only massive VP states therefore have the shape of a step function with a further decrease in frequency, as it is clearly seen in the blue lines representing $\Re[\sigma_{xx}]$ and $\Re[\sigma_{yy}]$, while transitions from the chiral to a massive VP state are governed by the linear dependence of the DOS on frequency [see the $0\rightarrow 1$ and $-1\rightarrow 0$ transitions in Fig. \ref{fig:3dsigmaBx}(a)]. Notice furthermore that the $0\rightarrow 0$ transitions, visible in $\sigma_{xx}$ and $\sigma_{yy}$, are special: the linear dependence of the joint DOS in frequency is canceled in the conductivity, which retrieves a universal value 
\begin{equation}
 \sigma_c=\frac{\pi}{8} \frac{e^2}{h}
\end{equation}
per Dirac cone, as it is known for graphene where the frequency-independent absorption is given in terms of the fine-structure constant $\alpha\simeq 1/137$ of quantum electrodynamics \cite{ludwig1994integer,nair2008fine}.

If we now take into account the in-plane magnetic field, we notice first that the transition thresholds for the surface-state transitions (orange lines in Fig. \ref{fig:3dsigmaBx}) are shifted to larger energies. This is expected from the globally reduced effective magnetic length $\gamma$, as mentioned above. One also notices the admixture of $n\rightarrow n\pm 1$ transitions in $\sigma_{yy}$, as expected from the selection rules. Furthermore, Landau quantization of the bulk bands yields effective (1D) bands with a parabolic dispersion at the band bottom [see Eq. (\ref{eq:bulkbands})]. This leads to a characteristic divergence in the (joint) DOS $\propto 1/\sqrt{\hbar\omega - \Delta_n}$, for $\hbar\omega\geq \Delta_n$, where $\Delta_n=\hbar v (\sqrt{2n/l_B^2 +\xi^{-2}}+\sqrt{2(n+1)/l_B^2+\xi^{-2}}) $ is the energy of the $n$-th bulk transition (purple lines in Fig. \ref{fig:3dsigmaBx}). 

       	\begin{figure}[h]
            \centering
            \subfigure{(a)\label{fig:3dsigmazzBz}
            \includegraphics[width= 0.45 \textwidth]{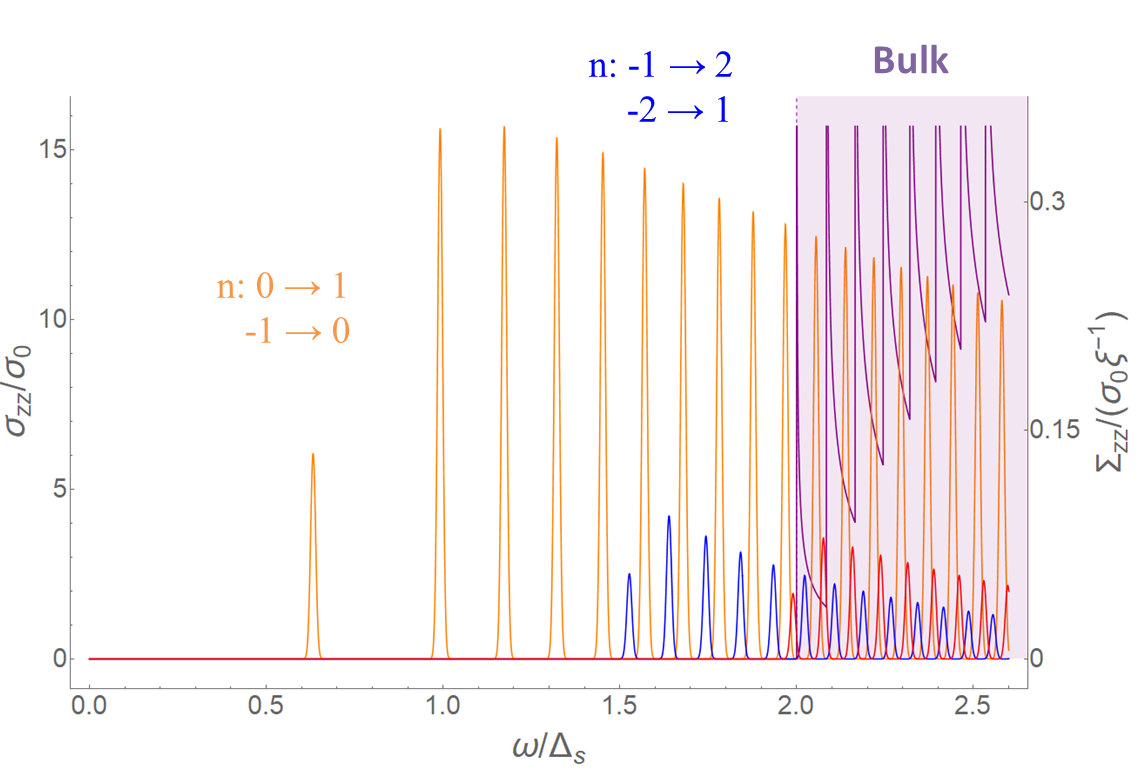}
            }
            \subfigure{(b)\label{fig:3dsigmaxxBz}
            \includegraphics[width= 0.45 \textwidth]{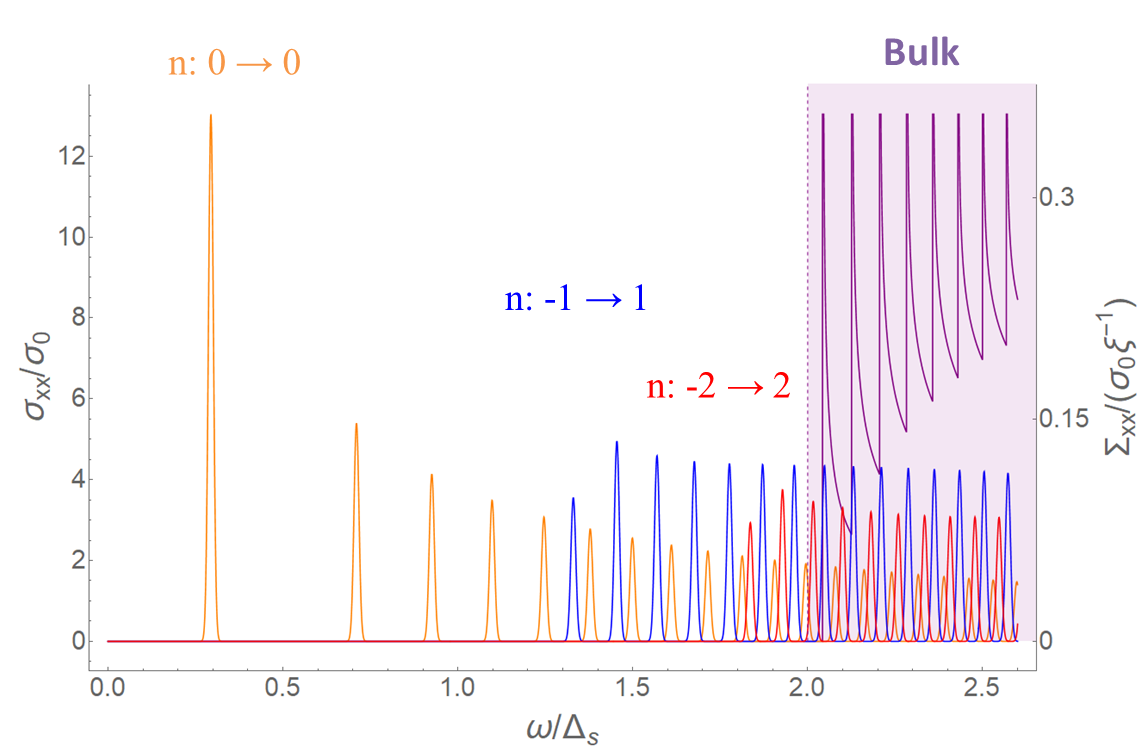}
            }
            \caption{Optical conductivities $\Re[\sigma_{zz}]$ and $\Re[\sigma_{xx}]=\Re[\sigma_{yy}]$, in units of $\sigma_0 =e^2/h$,
            in the presence of a magnetic field perpendicular to the interface. The magnetic field is chosen such that 
            $l_B = 2.15 l_s$, and excitation frequency is given in units of $\Delta_s$. We make Dirac distributions visible 
            by replacing them by a Gaussian distribution of standard deviation $a=0.01$ which can be seen as effect of finite temperature or disorders. 
            Different groups of peaks associated with the surface-band transitions $n$ are represented by different colors. Bulk transitions are plotted
            in purple, and the shaded region indicates the onset of the bulk gap.}
            \label{fig:3dsigmaBz}
            \end{figure}

To complete our analysis, we consider a magnetic field in the direction perpendicular to the interface, in which case the optical conductivites are plotted in Fig. \ref{fig:3dsigmaBz}. Due to rotation symmetry around the $z$-axis, we only need to consider $\sigma_{zz}(\omega)$ and $\sigma_{xx}(\omega)$ since $\sigma_{yy}(\omega)=\sigma_{xx}(\omega)$. Figure \ref{fig:3dsigmaBz} shows that the surface-state transitions are now fully quantized such that they are visible in the shape of peaks centered around the transition energies, while the bulk transitions maintain their asymmetric shape due to the 1D character of the Landau bands. The transition peaks can be regrouped into families labeled by the index $n$ associated with the confinement quantization if one considers the rather natural limit $l_s\ll l_B$, for small magnetic fields. Indeed, one may expect interface widths and values of $l_s$ in the nm range \cite{tchoumakov2017volkov}, while the magnetic length is typically on the order of 10 nm (for $B\sim 10$ T). The selection rules derived above are again clearly visible in the plots. 

In conclusion, we have performed calculations for the optical conductivity of a THJ that consists of a smooth interface between a TI and a trivial insulator. In addition to the chiral states, we retrieve massive VP states which exist as long as the interface width is larger than the intrinsic length $\xi =\hbar v/\Delta_s$. We have shown that these VP states have a distinct signature in magneto-optical spectroscopy as compared to the bulk transitions. First, they exist in an energy window below the bulk gap of the TI, and second the shape of their transition lines is distinct from the bulk transitions due to their different effective dimensionality -- while the bulk states become effective 1D bands upon the application of a magnetic field, the surface states maintain their 2D character. Most saliently, the surface states depend crucially on the orientation of the magnetic field. A magnetic field in the interface simply changes the position of the peaks and slightly modifies the optical selection rules, whereas a magnetic field perpendicular to the interface fully quantizes the surface bands and thus yields transition peaks. The orientation of the magnetic field is therefore an excellent probe of both chiral and massive surface states, as compared to the bulk states, which barely depend on the orientation of the magnetic field apart from a possible anisotropy of their band dispersion, which does not affect the characteristic shape of their transition peaks. 

We would like to acknowledge fruitful discussions with Sergue\"i Tchoumakov, David Carpentier, and Milan Orlita. This work was financially supported by the ANR project
``DIRAC3D'' under Grant No. ANR-17-CE30-0023.


%

\end{document}